\documentclass[useAMS,usenatbib,letterpaper]{mn2e}   % for astro-ph

\usepackage{graphicx}
\usepackage{natbib}
\usepackage{aas_macros}
\title[Photometric variability of TW Hya]
{Photometric variability of the T~Tauri star TW~Hya
on time scales of hours to years\thanks{Based on data from 
the MOST satellite, a Canadian Space Agency mission jointly 
operated by Dynacon Inc., the University of Toronto Institute for 
Aerospace Studies and the University of 
British Columbia, with the assistance of the University 
of Vienna, and on data from the All Sky Automated Survey (ASAS)
conducted by the Warsaw University Observatory, Warsaw, Poland 
at the Las Campanas Observatory, Chile.
}
}
\author[Slavek M. Rucinski et al.]
{Slavek M.\ Rucinski$^1$, 
Jaymie M.\ Matthews$^2$, 
Rainer Kuschnig$^2$, 
Grzegorz 
\newauthor
Pojma\'{n}ski$^3$, 
Jason Rowe$^4$, 
David B.\ Guenther$^5$, 
Anthony F.\ J.\ Moffat$^6$, 
\newauthor
Dimitar Sasselov$^7$, 
Gordon A.\ H.\ Walker$^2$, 
Werner W.\ Weiss$^8$ \\
$^1$Department of Astronomy and Astrophysics, 
University of Toronto, 50 St.\ George St., Toronto, 
Ontario, M5S~3H4, Canada\\
$^2$Department of Physics \& Astronomy, University of 
British Columbia, 6224 Agricultural Road, \\  
Vancouver, B.C., V6T~1Z1, Canada\\
$^3$Warsaw University Astronomical Observatory, 
Al.\ Ujazdowskie 4, 00--478 Warszawa, Poland\\
$^4$NASA-Ames Research Park, MS-244-30 Building 244, 
Rm 107A Moffett Field, CA 94035-1000, USA\\
$^5$Institute for Computational Astrophysics, 
Department of Astronomy and Physics, 
Saint Marys University, \\  Halifax, N.S., B3H~3C3, Canada\\
$^6$D\'{e}partment de Physique, Universit\'{e} 
de Montr\'{e}al, C.P.6128, Succursale: Centre-Ville, 
Montr\'{e}al, QC, H3C~3J7, and\\
Centre de Researche en Astrophysique du Qu\'{e}bec,
Canada\\
$^7$Harvard-Smithsonian Center for Astrophysics, 
60 Garden Street, Cambridge, MA 02138, USA\\
$^8$Institut f\"{u}r Astronomie, Universit\"{a}t Wien, 
T\"{u}rkenschanzstrasse 17, A-1180 Wien, Austria
}

\date{Accepted --.      Received -- ;      in original form --}

\pubyear{2008}

\begin{document}
\label{firstpage}
\maketitle

\begin{abstract}
MOST (Microvariability \& Oscillations of STars) and
ASAS (All Sky Automated Survey) observations have been 
used to characterize photometric variability of TW~Hya
on time scales from a fraction of a day to 7.5 weeks
and from a few days to 8 years, respectively. The
two data sets have very different uncertainties and 
temporal coverage properties and cannot be directly
combined, nevertheless, they suggests a global 
variability spectrum with ``flicker noise''
properties, i.e.\ with amplitudes $a \propto 1/\sqrt{f}$, 
over $>4$ decades in frequency, in the range
$f = 0.0003$ to 10 cycles per day (c/d).
A 3.7 d period is clearly present in the 
continuous 11 day, 0.07 d time resolution, 
observations by MOST in 2007. Brightness extrema
coincide with zero-velocity crossings in periodic (3.56 d) 
radial velocity variability detected in  
contemporaneous spectroscopic
observations of \citet{Set2008} and interpreted 
as caused by a planet. The 3.56/3.7~d periodicity 
was entirely absent in the second, four times longer 
MOST run in 2008, casting doubt
on the planetary explanation. Instead, a spectrum
of unstable single periods within the range of 2 -- 9 days 
was observed; the tendency of the periods to progressively shorten
was well traced using the wavelet analysis. 
The evolving periodicities and the overall 
flicker-noise characteristics of 
the TW~Hya variability suggest a combination of several
mechanisms, with the dominant ones 
probably related to the accretion processes
from the disk around the star.
\end{abstract}


\section{Introduction}
\label{intro}

Photometric variability of the very young, T~Tauri-type 
stars is still puzzling and remains an object 
of very active research; for the most recent
literature, see \citet{Percy2006} and \citet{ROTOR-I}. 
The variability comes in part from accretion and matter ejection
phenomena, in part from photospheric spots coming and
going into view, and in part from 
the inner accretion disk and then the accretion region 
as matter is channelled by magnetic fields into the
photosphere. 
\citet{Herbst1994} identified 3 basic types of 
variability which may coexist in a given 
T~Tauri star: Type~I, photospheric dark spots;
Type~II, variable accretion with some rotation modulation 
component; Type~III, totally random variations, 
mostly due to variable obscuration. 
Quasi-periodic variations, intertwined with chaotic 
changes are the most natural outcome of several
different mechanisms contributing simultaneously. 

Typical time scales of T~Tauri stars 
are of the order of hours to days,
but they are very difficult to characterize from the 
ground because of diurnal observation breaks 
and discontinuous temporal coverage. 
We are not aware of any attempt to obtain a
continuous record of T~Tauri variability from a satellite 
or through inter-observatory coordination of efforts. 

In this paper we present new, single broad-band 
(located between $V$ and $R$ bands), continuous 
photometric observations of the T~Tauri star 
TW~Hya obtained by the MOST satellite
mission over 11 days in 2007 and 46 days in 2008. 
Although the high frequency coverage (above
10 cycles per day) was inadequate, 
the temporal coverage at the once-per-orbit satellite
sampling period of 101.4 minutes was practically uninterrupted
permitting a study of the stellar variability 
on time scales of a fraction of a day to a few tens of days.
For still longer time scales, we used
the ASAS project data obtained over 8 yearly seasons from
2001 to 2008.
These data sampled the brightness changes of TW~Hya in the
$V$ and $I$ bands at intervals of a day to a few days. 

In this paper, after a brief introduction of TW~Hya itself
(Section~\ref{target}) and of its previous 
photometric variability studies (Section~\ref{prev}),
we discuss the results of the analysis of the 2007 and 2008
MOST data (Sections~\ref{most7} and \ref{most8}).
The TW~Hya variability at very low frequencies
is analyzed on the basis of the ASAS data
(Section~\ref{asas}). We conclude 
(Sections~\ref{disc} and \ref{concl}) that TW~Hya
shows a flicker-noise variability spectrum (the special type of 
a ``red noise'' spectrum) over a wide range of the 
time scale from hours to years.

\section{The target, TW Hya}
\label{target}

Following the suggestion of \citet{Herbig1978}, 
\citet{RK1983} established that TW~Hya (J2000:
11:01:51.9, $-$34:42:17) is a genuine,
if isolated T~Tauri star. 
Its spectral type is K7V and the spectrum is 
typical for the class with strong hydrogen line 
emission, with the Li 6707 \AA\ line present and 
with a complex and apparently irregular photometric 
variability. The spatial isolation of TW~Hya was puzzling and 
hard to explain. It appears on an empty sky field, 
far from any regions of star formation or other 
groupings of T~Tauri stars.
Later, however, an intense effort started by 
\citet{Reza1989} and continuing through several 
subsequent studies (e.g.\ \citet{Webb1999},
\citet{Zuckerman2001})
has led to a realization that TW~Hya is part of a loose
and dispersed association of young, nearby stars. But it
is the only one of two stars 
(the other is TWA~3A = Hen 3-600A; \citet{RayJay2006}) 
among them which continues to show disk accretion; the
remaining stars, members of what is now called the
TW~Hya Association (TWA), have properties of 
post-T~Tauri stars, i.e.\ 
still show a high abundance of lithium,
rapid rotation and resulting spot activity, but no
direct signs of disk accretion. 
While the TWA has about 25 definite members, \citet{Song2003}
broadened the definition of the association and added
more young members from other groups so that the current
count is about 45 stars. 

In this paper we limit ourselves to a study of the
photometric variability of TW~Hya, treating it as a
typical T~Tauri star. In fact, it is the nearest star 
of this type. Analysis of  raw
Hipparcos data \citep{Wichmann1998} determined
a moderately accurate parallax of 
$17.72 \pm 2.21$ mas, corresponding to a distance of 
$56 \pm 7$ pc, putting TW~Hya some twice as close as 
any other T~Tauri star. New reductions
of the Hipparcos data \citep{newhipp} gave a parallax
of $18.04 \pm 3.08$ mas ($55 \pm 9$ pc) confirming
the previous determination.

\citet{RK1983} observed variability of TW~Hya within 
$10.9 < V < 11.25$, $0.8 < B-V < 1.2$ and
$1.5 < V-I < 1.8$ and noted strong linear correlations
between these quantities (see Fig.~2 in that paper).
However, the {\it mean values\/} of the two
colour indices are not consistent: 
for the observed $V-I = 1.55$, an effective temperature
of about $4000 \,K$ would imply $B-V \simeq 1.3$, whereas 
$B-V \simeq 0.90$ is observed. 

Of interest to the interpretation of the variability of
TW~Hya is the low inclination of its rotational axis and
of its accretion disk. TW~Hya appears to
be visible nearly pole-on at a very low inclination
angle, most likely $i < 15^\circ$ \citep{Krist2000,Qi2004}.
\citet{Set2008} quote values of 
$i = 7^\circ \pm 1^\circ$ and $i = 14^\circ \pm 4^\circ$. 
\citet{Torres2003} found $V \sin i = 4$ km~s$^{-1}$ and cited 
several previous estimates 
of $V \sin i$: 4, 5, 10, 13, 14 and 15 km~s$^{-1}$.
\citet{RayJay2006} gave the new estimate of 
$V \sin i = 10.6$ km~s$^{-1}$. Because the rotation
period is likely to be of the order of 2 -- 4 days,
the low $V \sin i$ values imply a small inclination angle. 
The exact value is unimportant here;  the crucial point
is that the inner accretion disk is completely visible,
in contrast to many other T~Tauri stars which are 
-- in the majority -- detected with the normal probability 
of the inclination angle which scales as 
$\sin i$, i.e.\ usually they have large 
axial inclinations and are seen more edge-on.

% figure: MOST observations of 2007 --------------------------------
\begin{figure}
\begin{center}
\includegraphics[width=80mm]{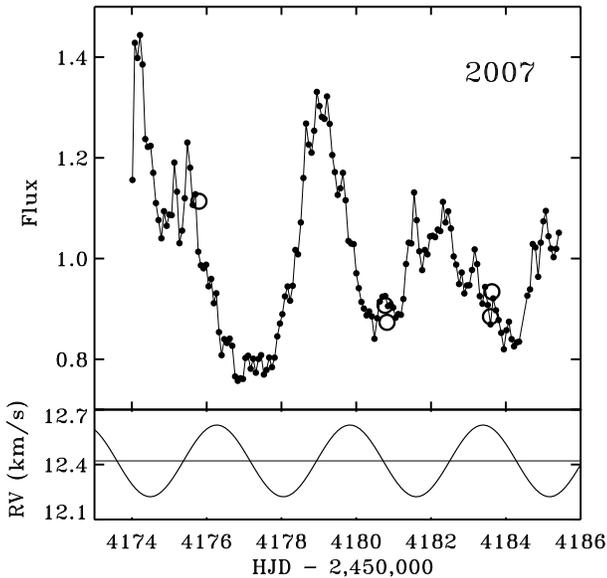}
\caption{The MOST data normalized to the average flux from the
11.4 days of observations of TW~Hya in 2007. The filled, connected, 
small circles give the mean flux values for each satellite 
orbit (the spacing is 0.0705 d; in two cases, twice
as long). The large circles show the
ASAS $V$ band observations, transformed to intensities, with the assumed 
mean value of $\bar{V} = 11.04$ (Section~\ref{asas}).
The lower panel gives the schematic radial velocity variations, 
based on the observations of \citet{Set2008} obtained just days before 
and after the 2007 MOST observations.
}
\label{MOSTdata7}
\end{center}
\end{figure}
%------------------------------------------------------------------------

\section{Previous observations of the photometric 
variability of TW Hya}
\label{prev}

TW~Hya shows a rich but confusing photometric
variability. \citet{RK1983} saw large night-to-night variations
of $\pm 0.2$ magnitude in $V$. From a few 
repeated nightly observations, they suggested rapid 
variability on time scales as short as 0.21 day. Later,
\citet{Rci1988} saw indications of a 2 day time-scale 
regularity. Several temporal variability investigations 
followed, as summarized in \citet{LC2005}. They indicated 
characteristic variability periods of 
2.88 days from Hipparcos photometric data \citep{KE2002},
2.85 days from $H\beta$ line-width variations and 
4.4 days from $B$ band veiling changes \citep{AB2002}.
\citet{LC2005} finally chose the period of 2.8 days as
the main, characteristic periodicity, but only after an arbitrary 
removal of about one fourth of their data
covering about 40\% of the whole duration of their
observing run. 

\citet{Herbst1994} suggested that TW~Hya
belongs to the Type~IIp of the T~Tauri variables
with the accretion variability combined
with a semi-periodic rotation modulation component.
The Type~IIp variables are characterized 
by periodic variations that persist for some time.

Recently, precise radial velocities of TW~Hya obtained by
\citet{Set2008} revealed a clear spectroscopic sinusoidal 
signal with a period of 3.56 d and amplitude of
$\pm 200$ m~s$^{-1}$. The authors interpreted it as an 
indication of a massive planet situated
inside the accretion disk, close to the surface of the
star, with orbital semi-major axis 
of only 0.041~AU = 8.8~R$_\odot$. 
To be sure that this is really the signal of 
planet-revolution, rather than that of the line-centroid
spot modulation reflecting star rotation, 
they re-analyzed (as summarized in the Supplementary 
material to this paper) all previous 
photometric data for TW~Hya. This 
gave several acceptable periods in the ranges of
1.74 -- 1.98 d, 1.98 -- 2.38 d, and 1.80 -- 2.20 d,
which were finally merged into an 
estimate of $2.1 \pm 0.5$ days. This period was identified
as the period of rotation of TW~Hya, as
distinct from the 3.56 d radial velocity modulation.  
Several other periodicities
were excluded in this process, but another one
surfaced, with a period of 9.05 days, as seen 
in the changes of the $H\alpha$ equivalent width.
We note that the planetary explanation of the radial
velocity changes has been recently questioned by
\citet{Huelamo2008} who propose photospheric spot-induced
spectral line shifts instead. 

Clearly, the picture of 
TW~Hya temporal variability is a very complex one. 
All previously suggested periods,
2.0, 2.1, 2.8, 2.85, 2.85, 4.4, 9.05 days -- in addition
to the three ranges suggested by \citet{Set2008} -- require
confirmation, particularly in view of the new 
radial velocity discovery of the strong (but still
different) spectroscopic periodicity of 3.56 days.
We have been fortunate that -- through 
sheer coincidence -- the MOST satellite 
observed TW~Hya photometrically exactly during the time
when radial velocity observations of \citet{Set2008} 
were collected. The satellite run lasted 11 days and indeed 
led to detection of a well defined photometric period of 3.7 days;
for such a short run, this value is in fact consistent with 3.56 d.
Because of the significance of this result and 
because the 2007 MOST run was too short to define this
period  well, we re-observed TW~Hya in 2008 over a time span
4 times longer than in 2007. 
The unexpected and intriguing results
of both MOST runs form the main part of this paper.

% Figure: Fourier analysis 2007 ---------------------------------------
\begin{figure}
\begin{center}
\includegraphics[width=80mm]{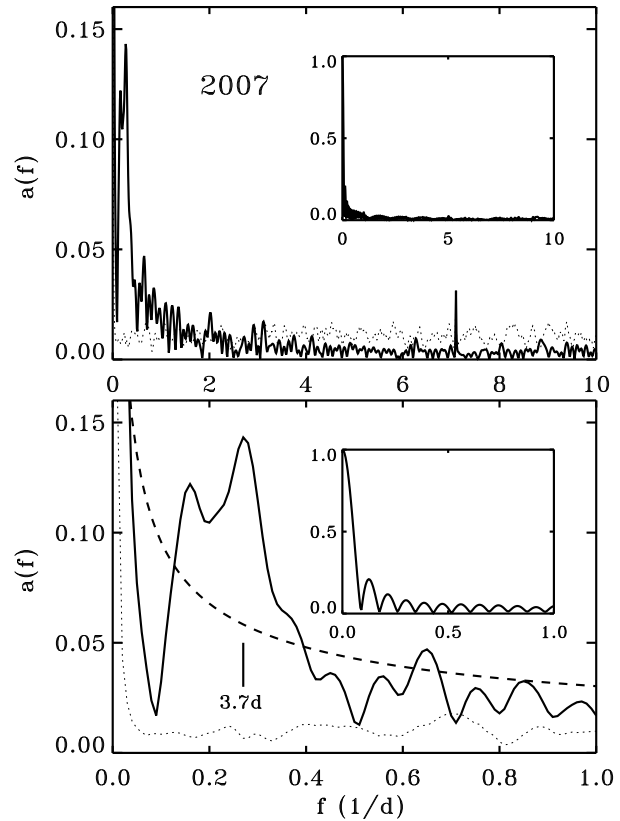}
\caption{The frequency spectrum expressed in
amplitudes for the MOST 2007 
data at frequencies below 10 c/d (the upper panel) 
and to 1 c/d (the lower panel).
The mean standard errors of the amplitudes, as
estimated by the bootstrap sampling technique,
are given by the thin, dotted line.
Note that the downward drift of the mean brightness level
through the observing run has resulted in a large
amplitude at the lowest frequencies.
The broken line shows the arbitrarily scaled
$a \propto 1/\sqrt{f}$ dependence. This line is
repeated in other similar figures later on, particularly
in Figure~\ref{MOSTfreq7_2} where it appears as the upper 
broken straight line.
Because of the practically uniform temporal 
sampling (only two data points missing in 163 consecutive
MOST orbits), the spectral window is exceptionally clean and
its side lobes are very small. 
}
\label{MOSTfreq7_1}
\end{center}
\end{figure}
%------------------------------------------------------------------------

\section{MOST 2007 observations}
\label{most7}

\subsection{The data}

MOST (Microvariability \& Oscillations of STars) is a microsatellite 
housing a 15-cm telescope which feeds a CCD photometer 
through a single custom, broadband, optical filter (350 - 700 nm). 
The effective wavelength of the single, broad-band filter 
is located between the $V$ and $R$ bands. The pre-launch 
characteristics of the mission are described by \citet{WM2003}
and the initial post-launch performance by \citet{M2004}.
MOST is in a Sun-synchronous polar orbit (820 km altitude) 
from which it can monitor some stars for as long as 2 months without 
interruption, provided they are located in the ecliptic part of 
the sky with declinations roughly in the
range $-18^\circ$ to $+35^\circ$.
The instrument was designed to obtain highly 
precise photometry of bright stars through Fabry-lens imaging, but 
direct imaging of fainter objects permits photometry in the
magnitude range of about 6 to 11 magnitude with typical 
accuracy of a few 0.001 mag. Since the loss of the
attitude control CCD system in 2006 and the need
to use the science CCD for the satellite stabilisation,
the data are obtained by stacking several short, 
1 -- 2 sec exposures into 
typically 30 sec data points. With the mean
$\bar{V} \simeq 11.0$ and $\bar{I} \simeq 9.3$
(see Section~\ref{asas}), TW~Hya is close to the faint 
limit of the MOST capabilities.   

In 2007, MOST observed TW~Hya for 11.4 days between 2007
March 14 and 2007 March 25. 
With declination $-34^\circ 42'$,
TW~Hya is located outside the Continuous Visibility Zone of 
the satellite. For that reason, the satellite had to be 
repeatedly re-oriented to a ``parking'' object
for part of its orbit (31~Com in 2007). 
This slightly affected the accuracy of the data, 
increased the instrumental noise which is 
due to stray light and South Atlantic Anomaly passages
and spoiled the shape of the Fourier spectral window.
In the analysis, for the sake of the uniformity of the data,
we sacrificed all variability information within each 
satellite orbit and formed single points separated on
the average by 101 minutes. 

For the 2007 observations, the median value of
the mean photometric error per one satellite-orbit point, 
resulting from averaging of 30 to 80 individual, 
1/2-minute exposures
spanning 15 to 40 minutes of time, was 0.0026
of the mean flux level. This number includes 
variability of the star within each satellite orbit.
The observed range of the mean errors was 0.001 to 0.005,
with a few observations with errors reaching 0.0075.

Variability of the star was certainly visible 
within each satellite orbit, but (1)~it was of a 
small amplitude, which is attested by small
values of mean standard errors formed from the
individual satellite orbits (see the third column
of Table~\ref{tab1}) and agrees with our main conclusions
on the dominant ``flicker noise'' variability (see further
in the paper) and (2)~aliasing due to the 
variable duration of the orbital scans
was severe. Thus, we analyzed only periods corresponding to 
frequencies of 13 cycles per day (c/d) or less; 
however, for simplicity, we have limited the analysis to
frequencies $f<10$ c/d.

% Figure: Fourier analysis log-log 2007 -----------------------

\begin{figure}
\begin{center}
\includegraphics[width=80mm]{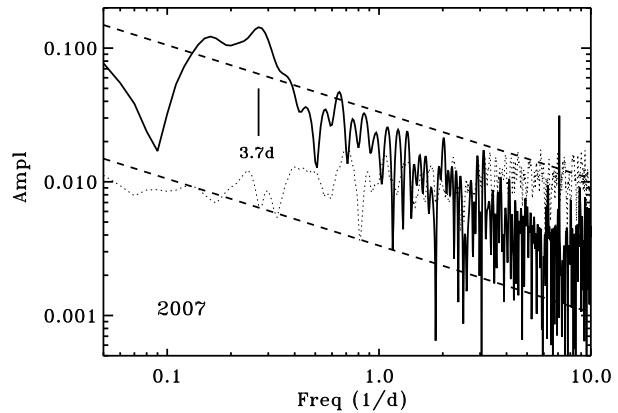}
\caption{The frequency spectrum expressed in
amplitudes for the MOST 2007 data 
of TW~Hya, plotted in log -- log units. The broken lines
give the slope of ``flicker'' noise, $a \propto 1/\sqrt{f}$,
whereas the thin, dotted line gives the approximate 
mean standard errors of the amplitudes.}
\label{MOSTfreq7_2}
\end{center}
\end{figure}
%--------------------------------------------------------------

% Table 1: data in ascii on-line for 2007 & 2008 MOST-----

\begin{table}
\begin{scriptsize}
\caption{TW~Hya: MOST 2007 and 2008 observations;
single data points per each satellite orbit. 
The whole table is available on-line only. \label{tab1}}
\begin{center}
\begin{tabular}{cccc} 
\hline
     (1)       &      (2)    &    (3)    &   (4) \\
     $t$       &   Mean flux &  Error    &   $n$ \\
\hline
   4174.0148   &   1.1560   &   0.0026   &   32  \\
   4174.0786   &   1.4284   &   0.0023   &   75  \\
   4174.1481   &   1.3983   &   0.0034   &   77  \\
   4174.2180   &   1.4437   &   0.0042   &   76  \\
   4174.2845   &   1.3854   &   0.0037   &   52  \\
\hline
\end{tabular}
\end{center}
\end{scriptsize}

\medskip
The columns: (1) $t  = JD - 2,450,000$;
(2) The photometric flux of TW~Hya for
each satellite orbit, independently
normalized to the mean value 
for the 2007 and 2008 observing runs; 
(3) The mean standard error of the normalized
flux estimated from the scatter of 30 -- 80
individual 0.5 minute integrations; 
(4) The number of individual observations 
contributing to the mean and used to evaluate the errors.
\end{table}
%-------------------------------------------------------------

The 2007 data are shown in Figure~\ref{MOSTdata7}
and are tabulated in Table~\ref{tab1}; this table 
contains also the MOST 2008 data discussed below.
The heliocentric time used in this paper is 
$t = JD - 2,450,000$.
% the ``MOST time'' $t_{MOST}$ counted from 
% January 1, 2000: $JD = t_{MOST} + 2,451,545$. 
The following features should be noted in 
Figure~\ref{MOSTdata7}: ($i$)~The light curve
consists of a gentle, but very well defined 
undulation with superimposed brightening events
lasting typically a fraction of a day, 
($ii$)~The slow variability shows 
three minima and four maxima indicating an
underlying period of  
about 3.5 days, ($iii$)~The overall variation appears to 
show larger amplitudes for longer time scales, a property
may be characteristic for some type of ``red noise''.

\subsection{Fourier analysis of the 2007 time series}
\label{FourierMOST7}

The Fourier analysis of the 2007 data was done by simple
least squares fits of expressions of the form
$l(f) = c_0(f) + c_1(f)\, \cos(2\pi (t-t_0) f) + 
    c_2(f)\, \sin(2\pi (t-t_0) f)$
for a range of frequencies $0.01 \le f \le 10$ c/d. For the 2007
data, the frequency step was $\Delta f = 0.01$. The
bootstrap sampling technique permitted evaluation of mean
standard errors of the amplitudes from the spread of the
coefficients $a_i$. This technique, for a uniform temporal
sampling -- as in our case -- may give too pessimistic 
estimates of errors (this seems to be actually 
the case, as seen in Figure~\ref{MOSTfreq7_2}), 
but we prefer this conservative approach. 
The amplitude $a(f)$ for each frequency was 
evaluated as the modulus of the periodic component,
$a(f) = \sqrt{{c_1(f)}^2+{c_2(f)}^2}$. Because of
the continuing changes in the spectrum (see further in the
paper), the phase information was disregarded except for
a comparison with the contemporaneous radial velocity
observations, as described below in Section~\ref{same-thing}

The amplitude spectrum is shown in Figure~\ref{MOSTfreq7_1}.
In this spectrum, all components with
frequencies $<2$ c/d (periods longer than half a day) 
appear to be significant and real. 
In agreement with what was noted in Figure~\ref{MOSTdata7},
we see an obvious periodic signal at $0.27 \pm 0.007$ c/d which 
corresponds to a period of $3.7 \pm 0.1$ d. Because the
data series was only 11 days long, we could not establish the
period more accurately, but it is consistent
with the periodic signal of 3.56 days in the radial velocity data,
as discovered by \citet{Set2008}. Because of the limited
duration of the run, we also consider the next peak in the
frequency spectrum corresponding to a 6.2 d period as 
non-physical. It should be noted that
no variability appears with any of the several 
periods listed in Section~\ref{prev}; in particular, 
the period close to 2.1 days, which was suggested as the stellar
rotation time scale by \citet{Set2008} is not visible at all.

% Figure: MOST observations 2008 ------------------------------
\begin{figure*}
\includegraphics[width=100mm,angle=90]{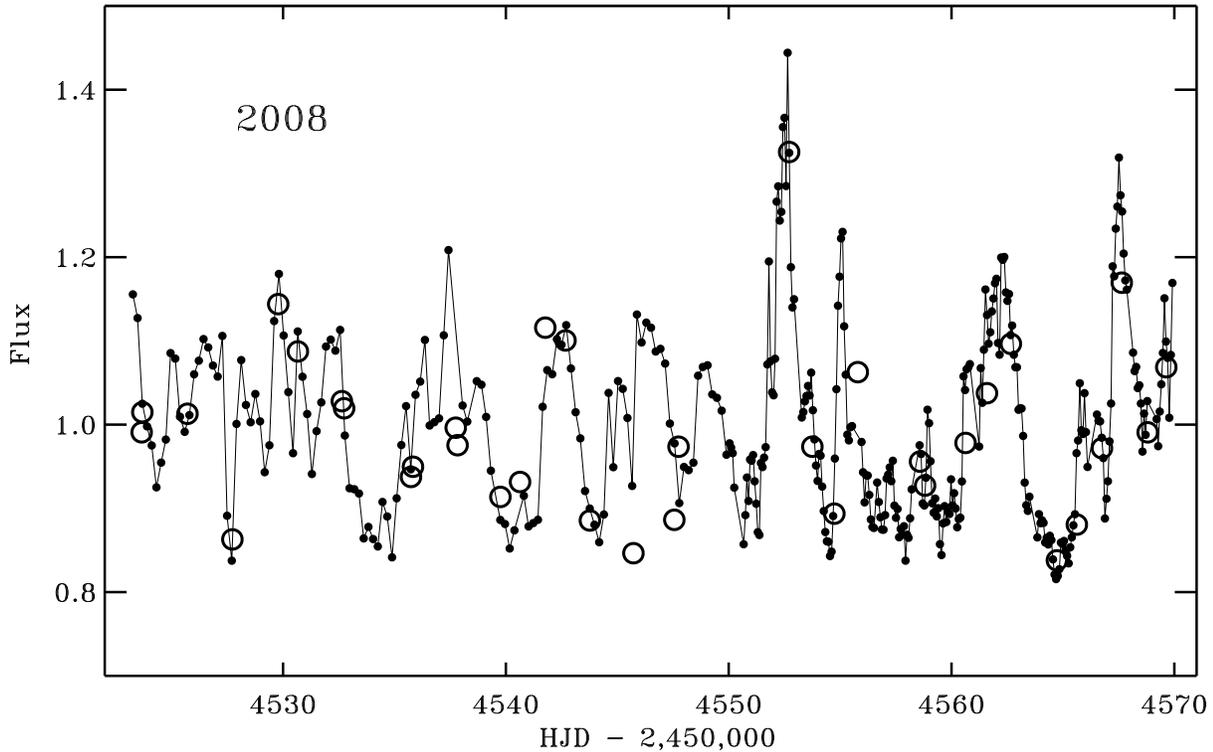}
\caption{MOST observations of TW~Hya in 2008 
with observations binned into mean points, 
separated by the intervals of (small) multiples of the
satellite orbital period of 0.0705 day.
Until day 4550, the data points are separated by
typically 2 -- 3 satellite orbits; after that, the sampling
became denser at every 1 -- 2 satellite orbits 
with regular (every 2.6 d), widely spaced gaps of 4 -- 5 orbits.  
Note that the Y-scale of this figure is the same as in
Figure~\ref{MOSTdata7}, although the 2007 and 2008 mean flux
levels were separately normalized and cannot be directly
compared. Note also that the transformed to intensities,
ASAS $V$ band observations (large, open circles) are plotted 
here with the same assumed reference
level $\bar{V} = 11.04$; see Section~\ref{asas}).}
\label{MOSTdata8}
\end{figure*}
%-------------------------------------------------------------

In Figure~\ref{MOSTfreq7_2}, we show the same amplitude
spectrum as in Figure~\ref{MOSTfreq7_1}, 
but in the log--log units. The spectrum clearly
rises at low frequencies in a way which is
characteristic for ``red noise''. 
As described lucidly by \citet{Press1978}, the special type
of red noise, called ``flicker noise'', with
amplitude spectrum $a(f) \propto 1/\sqrt{f}$
(i.e.\ the power $\propto 1/f$)
is very common and appears in various circumstances, 
although it is not clear why it is so prevalent in 
nature. Although, strictly speaking, the variability power 
diverges at low frequencies and is
non-integrable, there usually exists
a low frequency limit set by the slowest permissible 
response of the given dynamical system. 

The strong 3.7 day signal in the 2007 MOST data
appears to define the lowest-frequency periodicity, 
although we note
that the overall brightness of the star drifted down
during the span of 11 days. The 3.7 day periodicity
was stronger than estimated from a simple extrapolation of the 
flicker noise from the range of moderately
high frequencies of 0.7 -- 5 c/d into the low frequency end,
to $<0.5$ c/d.

\subsection{The 3.7 day photometric and the 3.56 day 
spectroscopic periodicities: The same thing?}
\label{same-thing}

Attempts at an analysis of the 2007 data using wavelets and
fractal techniques (see the description for the 2008 data, 
Section~\ref{wave8}) showed that the 2007 run was 
simply too short to state anything beyond 
the existence of the very clear 3.7 day periodicity, 
superimposed on (or as part of) 
a more complex flicker-noise variability. 
In fact, in view of this particular type of 
variability, even the 3.7 day periodicity may be considered
questionable in view of the remark of
\cite{Press1978} that any period equal to 1/3 of the length
of the flicker-noise dominated data is probably spurious. 
If not for the presence of the 3.56 day periodicity 
observed at the exactly same time,  
in the very differently obtained 
radial-velocity data of \citet{Set2008}, one
would be tempted to interpret the 3.7 day period
as an exceptionally large flicker-noise ``fluke''. 

The 2007 MOST run was located between the two sections
of the \citet{Set2008} radial velocity observations
which had a gap of 1.5 months between $HJD - 2,450,000$ days
4172 and 4215. Apparently, the RV variations kept 
the same phase through the duration
of this gap so that one can plot the
interpolated RV changes for the dates of the MOST
observations. To do that, 
the phase information for the \citet{Set2008} data has been
restored from the original RV observations and the expected
variations plotted in the lower panel of 
Figure~\ref{MOSTdata7}. 
The results are very important for the interpretation 
of the 3.56/3.7 day periodicity: The photometric
and spectroscopic observations were synchronized in 
such a way that the
largest RV excursions occurred at the time 
of the fastest brightness changes, i.e.\ the two types of
variation were shifted in phase 
by 90 degrees. Thus, for the orbital motion
of a planet or of a gas blob inside the disk,
the photometric extreme values were reached 
during what (in binary star language) would be 
called the ``conjunctions'': The star was brightest when the
RV went through the zero deviation, when switching for
the approach to the recession part of the cycle 
(the planet/blob in front the star), 
and was faintest when the RV went through 
the zero deviation from the recession to the approach
(the planet/blob behind). This type of the phase
relation would agree with the spot explanation, but
-- if the star is really seen pole-on -- the
spot modulation is expected to be small and could not lead
to the observed brightness variations by 30 -- 40\%.

\section{MOST 2008 observations}
\label{most8}

\subsection{The data}

MOST observed TW~Hya the second time for 46.7 days, 
from February 26 to April 13, 2008. 
All remaining details of the observations were the
same as for the 2007 run except that: 
(i)~the switch targets were different, HD~99563 during the
first month and HD~102195 for the last 17 days;
(ii)~the star was observed typically every second or 
third MOST orbit; 
(iii)~during the period of March 24 to April 12, the TW~Hya 
observations were interrupted for 5 -- 7 satellite orbits 
every 2.6 days.
These restrictions resulted in larger gaps 
in observations than in 2007, 
but the gaps did not affect the low variability 
frequencies which are the main target of our interest. 

The 2008 data are listed in Table~\ref{tab1}
and are shown in Figure~\ref{MOSTdata8}. Note that the 2007
and 2008 fluxes have been separately normalized because 
of (possible) small differences
in the satellite sensitivity precluded direct
ties of the 2007 and 2008 seasonal mean levels.  
However, the ground-based $V$ band data do not indicate
any large change in the mean light level; 
see Section~\ref{asas} where the variability of TW~Hya 
over time scales of months and years is analyzed.

The individual MOST-orbit observations lasted between
12 minutes and 33 minutes and were typically spaced by 
small multiples of the
satellite orbital period of 101.4 minutes.
The formal errors per mean satellite-orbit
point have the median of 0.0018
of the mean flux level and the range of 
0.001 to 0.005, with a small number of relatively 
poorer observations having uncertainties 
approaching 0.007.

\begin{figure}
\begin{center}
\includegraphics[width=70mm]{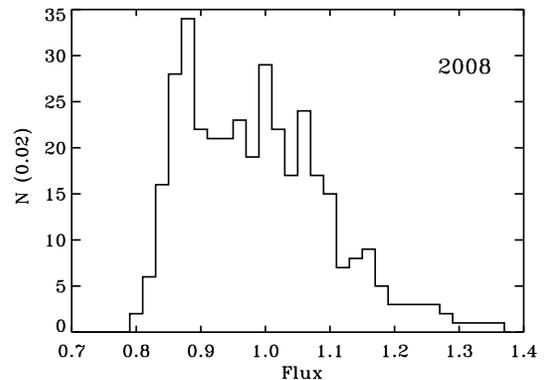}
\caption{The distribution of deviations from the mean
flux level for the MOST 2008 data indicating the presence of
relatively short lasting spikes superimposed on a moderately
steady background level.} 
\label{MOSThist8}
\end{center}
\end{figure}

The general characteristics of the 2008 TW~Hya variability are the 
same as in the 2007 observations, but we clearly see and can
follow much more activity in such a long data series. 
Brightness increases appear to be more common than 
dimming events. This is visible directly in 
Figure~\ref{MOSTdata8} and through the 
skewness (+0.772) of the distribution of the deviations
from the mean level (Figure~\ref{MOSThist8}),
where upward spikes produce a positive tail in the 
distribution\footnote{We have not performed this exercise
for the 2007 data because this would require to remove
the slow changes which do not have a clear interpretation
or description.}.

% Figure: MOST 2008, distribution of deviations ---------------

%--------------------------------------------------------------

% Figure: Fourier analysis 2008 -------------------------------

\begin{figure}
\begin{center}
\includegraphics[width=80mm]{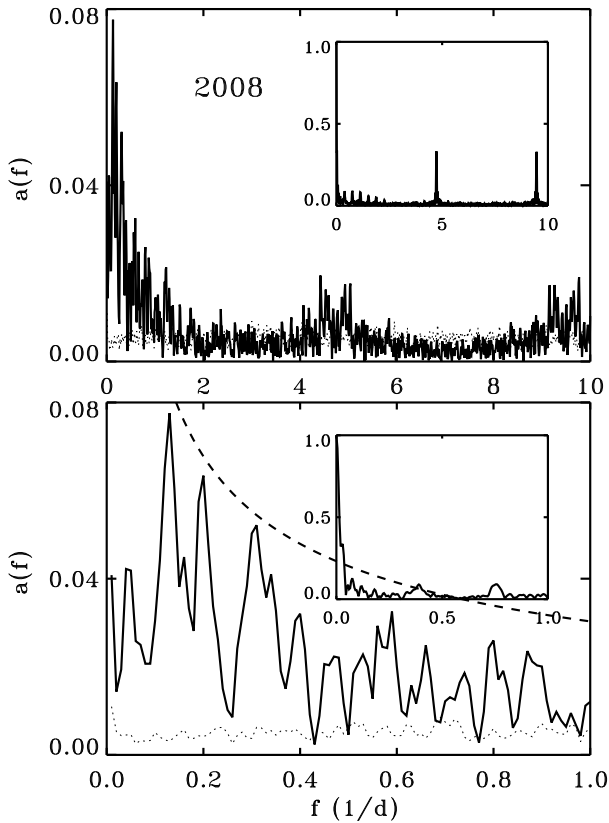}
\caption{The frequency spectrum expressed in
amplitudes for the MOST 2008 data, for frequencies up
to 10 c/d (the upper panel) and up to 1 c/d (the lower panel).
Compare this with Figure~\ref{MOSTfreq7_1} and note
that the amplitudes are smaller than in 2007 so that  
the vertical scale is reduced here by a factor of 2.
The broken line shows the arbitrarily scaled
$1/\sqrt{f}$ dependence; it is the same upper envelope as
in Figures~\ref{MOSTfreq7_1}, \ref{MOSTfreq7_2} 
and \ref{MOSTfreq8_2}.
The inserts show the spectral window. Note that it is
not as ``clean'' as for the 2007 observations because only
364 of 663 consecutive MOST orbits were used, with the most
common spacing of 3 satellite orbits. 
}
\label{MOSTfreq8_1}
\end{center}
\end{figure}
%--------------------------------------------------------------

% Figure: Fourier analysis 2008 log-log -----------------------

\begin{figure}
\begin{center}
\includegraphics[width=80mm]{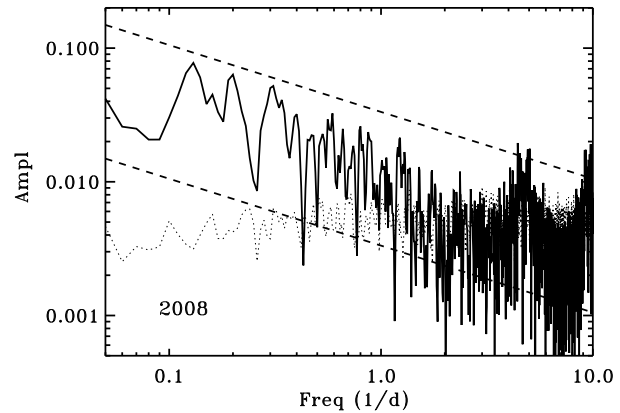}
\caption{The frequency spectrum expressed in
amplitudes for the 2008 MOST data 
of TW~Hya plotted in log -- log units. The broken lines
give the slope of the ``flicker'' noise, $a \propto 1/\sqrt{f}$,
whereas the thin, dotted line gives the approximate 
mean standard errors of the amplitudes. The vertical scale is 
the same as in Figure~\ref{MOSTfreq7_2} so one can see here
the absence of the strong 3.7 day periodicity.}
\label{MOSTfreq8_2}
\end{center}
\end{figure}
% --------------------------------------------------------------

\subsection{Fourier analysis of the 2008 time series}
\label{FourierMOST8}

The Fourier analysis of the 2008 data was done
exactly in the same way as for the 2007 data. The same
relatively dense frequency sampling of 
$\Delta f = 0.01$ was used which 
was more appropriate than in 2007 in view of the longer
duration of the run.

The picture (Figure~\ref{MOSTfreq8_1})
is now very different from that in 2007: The strong
3.7 d periodicity, so well defined in 
the 2007 run, is entirely absent.  
Instead, a number of periodic components, 
with slightly smaller amplitudes than the 2007
periodicity, appear to be present at $f <1$ c/d;   
their formally derived periods are
7.7, 5.1, 3.3, 2.5, 1.74, 1.25 days. Note that all
are very highly significant as the comparison of
the amplitudes with their errors in 
Figure~\ref{MOSTfreq8_1} clearly shows.
Their amplitude progression indicates a modulation 
by an envelope $\propto 1/\sqrt{f}$, 
i.e.\ again pointing to flicker noise. The feature at
$f \simeq 4.5 - 5$ c/d (0.21 -- 0.22 day) appears
in the spectral window and 
is an artefact of the gaps in the data occurring
at the spacing of $3 \times$ the satellite period.
The spectral window is however relatively ``clean''
in the low frequency range ($0 < f < 1$ c/d) so
that all components visible in the lower panel of
Figure~\ref{MOSTfreq8_1} are real and well defined.

The log--log plot (Figure~\ref{MOSTfreq8_2}) 
confirms the flicker-noise amplitude distribution. The slope
seems to be locally slightly steeper in the range $0.15 < f < 1.5$
c/d (periods $\simeq 0.7 < P < 7$ days) as if slower
variations were a bit more likely to appear than 
for the strict flicker noise. Viewed in this light,
the 3.7 day variability observed in 2007 could be an extreme
manifestation of this tendency.

\subsection{Wavelet analysis of the 2008 time series}
\label{wave8}

The Fourier analysis presented in Sections~\ref{FourierMOST7}
and \ref{FourierMOST8} addresses the periodic content
in the TW~Hya variability. But, from the comparison of
the 2007 and 2008 data, we know that the periods of the variability 
must change in time. While Fourier analysis cannot trace
such temporal changes, the wavelet technique has been developed
specifically as a tool to localize in time finite 
wave trains of various periodicities and durations.
Extensive literature on the subject exists; of particular
use to us was \citet{TC1998} following the ideas developed in 
the seminal work of \citet{Daub1992}.

The wavelet and fractal (the next Section) analyses 
require the data to be spaced uniformly in time. 
To achieve this time uniformity, the 2008 MOST
data have been mapped into a strict equidistant grid of
points spaced at 0.07047 d (the satellite revolution period)
using splines. Deviations of the actual observations from
uniformity of the time scale were small, typically 
1 -- 2 minutes. However, because of the different
satellite target switching, gaps 
lasting typically two or three satellite orbits occurred. 
The spline interpolation into the uniform scale 
resulted in 663 points spanning the same time range as
the original data. 

% Figure: wavelet, grey scale --------------------------------

\begin{figure*}
\includegraphics[width=110mm,angle=90]{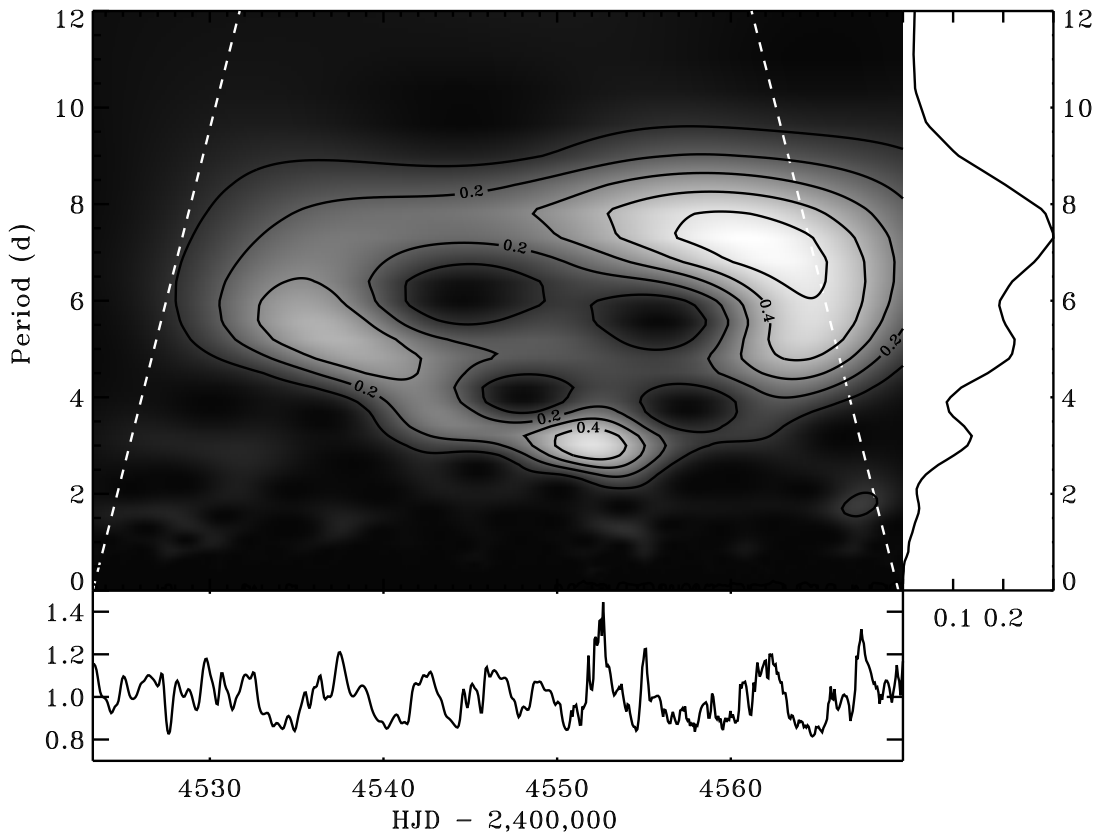}
\caption{The Morlet wavelet transform of the 2008 MOST 
data. The grey scale gives the power
of the transform. The edge effects are present beyond the
white, slanted, broken lines. The TW~Hya brightness data, 
re-sampled into a uniform grid with time-point spacing of 
0.07 d are shown at the bottom. The projected
mean frequency spectrum is shown at the right
margin; compare with Figure~\ref{MOSTfreq8_1},
but note the poorer definition of the spectrum here, the
price paid for localization of the periodic events.
}
\label{WL8}
\end{figure*}
%-------------------------------------------------------------

The 2008 data were subjected to wavelet analysis using
several different types of wavelet functions.
The best and most clearly defined results were
obtained with the simplest (sometimes considered
the ``natural'') Morlet, complex wavelet consisting of a 
coupled sine--cosine pair modulated by a Gaussian 
function.

The wavelet power (the squared modulus of the
transform) for the 2008 observations of TW~Hya is shown
in Figure~\ref{WL8}. We do not address the matter of
units of the power\footnote{We used the routine ``wv\_cwt''
in the IDL~6.3 software to calculate the power
of the wavelet components.}
and utilize only the periods and the time
localization of the periodic wave packets.  
The wavelet transform was calculated
with the usual, power-law time-scale
progression (1, 2, 4, 8,... data point
spacing) but for the ease of viewing and 
interpretation, the transform has been interpolated into a linear 
time scale. In the grey-scale image, a single bright spot
corresponds to a well defined periodic packet which
lasts about 5 -- 6 oscillation periods while 
any horizontal widening of it would indicate
a longer duration of the periodic wave.   

The results of the wavelet analysis of the TW~Hya MOST data
(Figure~\ref{WL8}) are striking and very important: 
Periodic oscillations apparently appeared
at some periods, lasted for some time and died out. 
During their lifetimes, they had a tendency to
shorten the period. In particular,
a periodic variation with a period of about 5 -- 6 days 
started around the day $t \simeq 4530 - 4533$ 
and rapidly shortened its period to about 3 days 
in some 10 days; possibly, the isolated
2.5 d periodicity appearing later around day 4553 
is actually an extension of
this progression. Another periodic variation started  
at about day 4550 -- 4552 with a period of about 7 d, 
became very strong at about the day 4560 
evolving into a wider, less
concentrated feature of a slightly shorter period;
it could not be followed
further because of the end of the data.

Although our analysis extended to both,
short ($P \ge 0.2$ d) as well as long ($P \le 20$ d)
periods, there is no indication of any periodic 
activity outside of the range of $2 < P < 9$ days. 
This very well agrees with the Fourier analyses
presented before and suggests a rather well defined 
range of temporal scales. At a given time usually
only one periodicity was present. The several 
spectral features in Figure~\ref{MOSTfreq8_1} 
with frequencies $>0.5$ c/d are most probably 
harmonic artefacts of single wave trains changing their
periods. As noted in the description
of the Fourier analysis (Section~\ref{FourierMOST8}),
we see no trace of any dominant periodicity with a period of
3.5 -- 3.7 days; thus, the strong signal in the 2007 data
was entirely absent in the 2008 observations.

A comment is necessary here about the assumed 
duration or the ``order'' of the Morlet wavelet
analyzing function. In the literature, the order
of the Morlet wavelet packet is frequently not explicitly
given, but appears to be usually assumed to describe 
five sine--cosine cycles per one Gaussian-enveloped
packet. An attempt was made to utilize the Morlet
wavelets of different orders to find the optimum fit to
the {\it duration of the wave packet, expressed 
in the number of periods\/}. Unfortunately, we
found a unexpected problem: In the wavelet analysis, there 
appears to exist a coupling between the duration of
the packet (the length of the Gaussian envelope)
and the oscillation period. In other
words, the packet of a given period will show
at slightly different period 
depending on the assumed order. As a result,
for different orders, the whole two-dimensional 
wavelet transform structure, as
in Figure~\ref{WL8}, is stretched or compressed
vertically depending on the order. 
We were able to remove this arbitrariness
by imposing the condition that the time-averaged power 
distribution (the right side of that figure) has the same
shape as the Fourier spectrum power. Using this principle, 
we found that Morlet-6, and not Morlet-5, 
is the most appropriate. 
Further tests on artificial data for the Morlet 
orders 4 -- 7 showed that a typical shift in 
the period scale is about 20\% per increment 
in the wavelet order. 
This property does not seem to be widely known 
or described but it appears to be a limitation of the wavelet 
approach. It precludes our original hopes of the evaluation
of how many cycles are confined in a typical wave packet
for TW~Hya. 

Very similar results, with similar time scales (periods) 
and wave-packet localizations as for Morlet-6 were
obtained for the 2008 data with another popular  
wavelet function, the Paul-4 wavelet. However, 
an application of the wavelet analysis 
to the 2007 data has not led to any
new results: The main 3.7 day single periodicity
dominated the picture; its amplitude
diminished through the 11 day MOST sequence so that
the formally derived, single-period wavelet had 
a maximum in the first part of the run.

% Figure: correlation integral --------------------------

\begin{figure}
\begin{center}
\includegraphics[width=80mm]{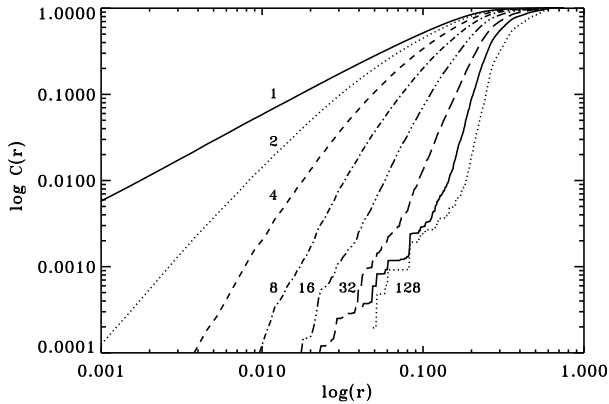}
\caption{The fractal properties of the 2008 MOST data 
of TW~Hya analyzed using the correlation
integral $C(r)$. The line labels (powers of 2)
give the lengths of segments used in the calculation
of $C(r)$; they can be converted into the time
scale lengths by multiplying them by the length
of the data spacing  of 0.07047 d.
For definitions and conventions, see the cited 
literature, in particular \citet{Lehto1993}.}
\label{MOSTfract_CI}
\end{center}
\end{figure}
%--------------------------------------------------------

\subsection{Fractal analysis of the 2008 time series}

An attempt to analyze the
irregular photometric variability of TW~Hya
for any indications of a deterministic process was made. 
The same 2008 time series as the one described above 
in the wavelet analysis and consisting of 663 
equidistant points was used here.
The fractal technique utilized 
the correlation integral, first described by \citet{GP1983}
and studied extensively in \citet{Voges1987} and 
\citet{Lehto1993}.
This approach may -- in some circumstances -- 
permit evaluation of the fractal dimension (if such
can be defined) and may reveal the type of variability.
The data are sampled in segments of progressively
larger length and ``distances'' are calculated between
all pairs of such segments; then the number of
pairs satisfying a criterion of the distance is found.
The reader is directed to the paper of \citet{Lehto1993}
for the description of the terms and definitions used in 
this technique.

In Figure~\ref{MOSTfract_CI} we see
that the formally derived dimension (from the 
logarithmic slope of the correlation 
integral $C(r)$ versus the distance of the points, $r$)
is not unique and increases with 
the size of the embedded dimension $d$
in the range of 1 to 32 data intervals, i.e.\ in the
time scales of 0.07 to 2.2 days.  
As discussed by \citet{Lehto1993} this is a property 
characteristic for independent, uncorrelated
shot-noise events. 
Beyond the point of the time scales  
corresponding to $\simeq 2$ days, the correlation
integral changes its character and a ragged structure in
$C(r)$ becomes visible for the time scales of 
2 -- 9 days. This structure appears to be  
related to the semi-regular variability so clearly 
manifested in the wavelet analysis. This result
fully confirms lack of any periodic events
in the wavelet analysis for time scales $<2$ days.

% Figure: ASAS data -----------------------------------------

%\begin{figure}
\begin{figure*}
\includegraphics[width=100mm,angle=90]{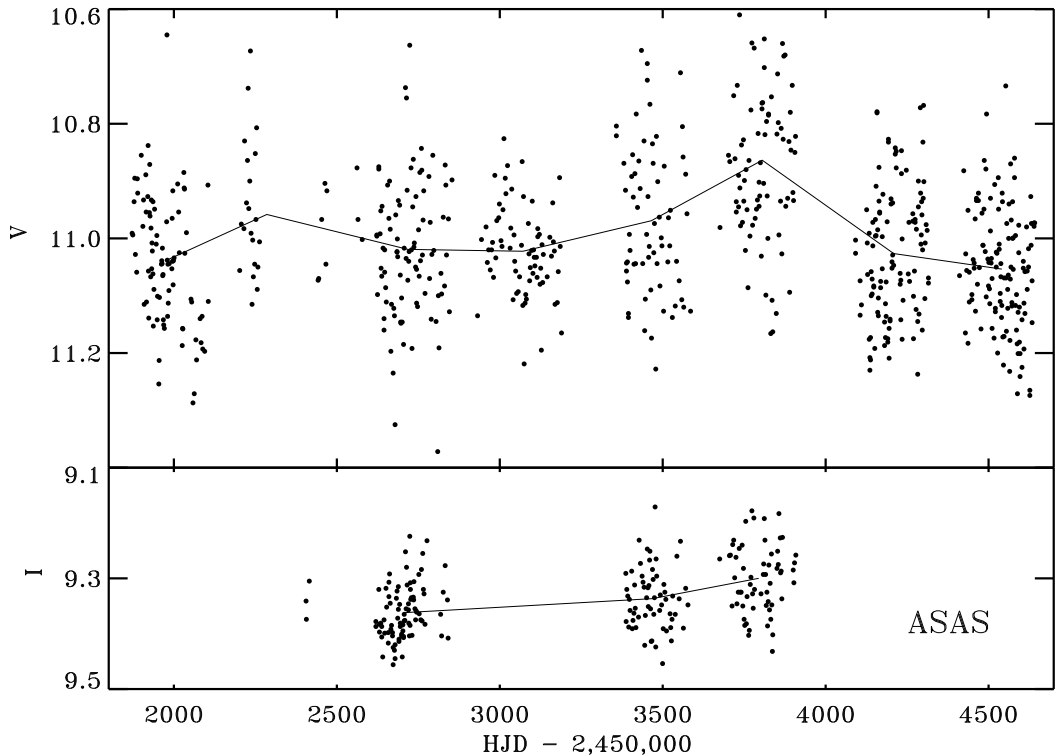}
\caption{The ASAS data for TW~Hya
in $V$ and $I$ band filters in eight consecutive
observing seasons, 2001 -- 2008. Most of the scatter
is due to variability of TW~Hya; the observational
errors in both filters are at the level of 0.01 -- 0.02
mag. The lines connect the mean seasonal values 
of the magnitudes.}
\label{ASASdata}
\end{figure*}
%\end{figure}
%-------------------------------------------------------------

\section{ASAS observations}
\label{asas}

\subsection{The ASAS data}

ASAS, the All Sky Automated 
Survey\citep{Poj1997,Poj2002,Poj2004,Poj2005,BP2006}\footnote{See:
http://www.astrouw.edu.pl/$\sim$gp/asas/asas.html and 
http://archive.princeton.edu/$\sim$asas/ }, is a long 
term project dedicated to detection and monitoring of
variability of bright stars using small telescopes. 
It has been run by the Warsaw University at the Las Campanas
Observatory in Chile using 7~cm telescopes providing
the best photometry in the 8 -- 13 magnitude range. 
About three fourths of the sky (the southern and equatorial
parts, with $-90^\circ < \delta < +28^\circ$)
have been monitored for stellar variability 
in $V$ and $I$ filters. The typical random errors are
at the level of 0.01 -- 0.02 mag. The high stability of the
system and of the photometric reductions and calibrations 
results in a very consistent data set.

The ASAS observations of TW~Hya analyzed here cover 
8 yearly seasons, 2001 -- 2008, centred on 
late February of each year. The photometric
observations are plotted versus the HJD in Figure~\ref{ASASdata}.
As was noticed before by \citet{RK1983}, (i)~the variability
in the $V$ band is much stronger than in the
$I$ band, (ii)~the $V-I$ colour 
index changes follow the $V$ changes so that the
$V$ band variations are larger and easier to study than 
the $I$ band ones. Note that with 
$\bar{V} \simeq 11.0$ and $\bar{I} \simeq 9.3$, 
TW~Hya is an easy object for ASAS.  
The measurement errors in both bands are similar,
at the level of 0.01 -- 0.02 mag, and reflect mostly 
the uncertainties in the standard system transformations
over large ASAS fields. 

While the night-to-night variability of TW~Hya 
appears as random scatter in Figure~\ref{ASASdata}, 
the seasonal data show well defined, slow changes with 
$\Delta V = 0.18$, from $V = 11.05$ in 2003 to 
$V = 10.87$ in 2006 and by $\Delta I = 0.06$,
from $I = 9.36$ in 2003 to $I = 9.30$ in 2006. 
Twenty years earlier, in 1982 \citep{RK1983}, the star was 
slightly fainter with mean $V \simeq 11.15$ 
and $I \simeq 9.5$, but the night-to-night variability 
ranges were similar, with $\Delta V \simeq 0.35$ and 
$\Delta I \simeq 0.15$.

% Table 2: data in ascii on-line for 2001 - 2008 ASAS  ---------

\begin{table}
\begin{scriptsize}
\caption{ASAS photometric data for TW~Hya, 2001 -- 2008. 
The whole table is available on-line only. \label{tab2}}
\begin{center}
\begin{tabular}{crrc} 
\hline
      (1)    &    (2) &  (3)   & (4) \\
  $JD - 2,450,000$ & $V/I$ & Error & Filter \\
\hline
   2405.5505 &  9.341 &  0.063 &  I  \\
   2406.5620 &  9.374 &  0.062 &  I  \\
   2415.5269 &  9.305 &  0.061 &  I  \\
   2619.7858 &  9.378 &  0.058 &  I  \\
   2619.8205 &  9.387 &  0.060 &  I  \\
\hline
\end{tabular}
\end{center}
\end{scriptsize}

\medskip
The columns: (1) The time, $t_{ASAS} = JD - 2,450,000$;
(2) The magnitude in the $V$ or $I$ filters, (3) An estimate of the
mean error of the magnitude provided by the ASAS
project; it should be taken in the relative 
sense, in comparison between observations and is usually
an over-estimate; (4) The filter used; in the tabulation,
first $I$ then $I$. 
\end{table}
%-------------------------------------------------------------

\subsection{Fourier analysis of the whole dataset}

Because the $I$ band variability of TW~Hya appears to follow 
the $V$ band variability but with reduced scale, we present here
periodic analysis of the $V$ data only. We should note that
the $V$ variability cannot be directly compared with 
that observed by MOST. The MOST single filter is located 
red ward of the $V$ band effective wavelength, so that
the $V$ band amplitudes of TW~Hya are expected to 
be slightly larger than the MOST amplitudes.

%Figure: ASAS window all seasons  ---------------------------

\begin{figure} 
\begin{center}
\includegraphics[width=75mm]{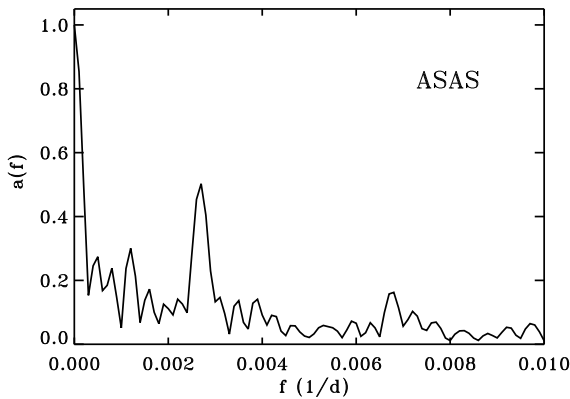}
\caption{The spectral window for all
$V$ band ASAS observations of TW~Hya during
the 2001 -- 2008 seasons. The window is free of any 
features at frequencies higher than shown in the figure.
}
\label{ASASwind}
\end{center}
\end{figure} 
%--------------------------------------------------------------

The whole span of the ASAS observations extends over about
3000 days setting a limit to
the lowest detectable frequencies of 0.0003 c/d.
We consider here the frequencies up to 0.5 c/d which
would be accessible for the one-day sampling. In fact,
the most frequent data spacing was two days; it occurred
for 43\% of the $V$ observations 
and 51\% of the $I$ observations. A smaller 
number of observations were spaced by one day, 
17\% of $V$ and 10\% of $I$ observations, and 
5\% and 7\% of all $V$ and $I$ observations 
were done twice on a given night. The resulting spectral
window is relatively simple (Figure~\ref{ASASwind}) and
shows only the yearly signal at 0.00274 c/d
and a feature at 0.0677 c/d corresponding
to the time scale of 148 d, somewhat similar to
the duration of each seasonal run.
 
\begin{figure} 
\begin{center}
\includegraphics[width=80mm]{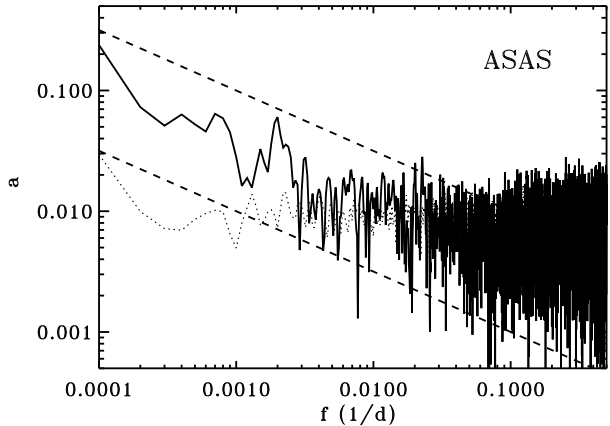}
\caption{The amplitude frequency spectrum for all
$V$ band ASAS data of TW~Hya of the 2001 -- 2008 seasons,
plotted in log--log units. The broken lines
give the slope of flicker noise, $a \propto 1/\sqrt{f}$,
%(with different normalizations from those used for
%the MOST spectra), 
whereas the thin, dotted line gives the approximate 
mean standard errors of the amplitudes.}
\label{ASASfreqLog}
\end{center}
\end{figure}

The frequency analysis of the ASAS data
was done in the same way as described
in Section~\ref{FourierMOST7}, through least squares
fits and bootstrap error estimates. The bootstrap technique
is particularly useful here because of the unequal 
temporal distribution of the data. 
In the analysis of the full $V$ band ASAS dataset, 
the seasonal trends were {\it not removed\/} to
retain the low frequency component of the variability.
While the spectrum  expressed in linear units is featureless,
indicating lack of coherence over 8 years, 
Figure~\ref{ASASfreqLog} shows that in the wide
range of frequencies of 0.0001 to 0.5 c/d, 
the spectrum approximately follows a
``flicker-noise'' dependence. The seasonal 
mean-level changes are apparently part of this picture.
Because of the long extent of the ASAS data, the
low frequencies are very well defined, in spite
of the ASAS $V$-amplitude errors at the level
of $\sigma_a \simeq 0.01 - 0.02$. At higher frequencies
accessible to MOST observations ($f > 0.02$ c/d),
the MOST results are far superior over
the ASAS ones, mostly because of the amplitude
errors by an order of magnitude smaller than for ASAS
(see Figure~\ref{MOSTfreq8_2}).

The maxima along the low-frequency ``flicker-noise''
progression in Figure~\ref{ASASfreqLog}, 
corresponding to periods of 155, 317 and 
510 days appear to be significant ($>3 \sigma$),
but their reality is not fully established. 
In the frequency range of $0.05 < f < 0.5$ c/d
(time scales 2 -- 20 d)
the observational noise with similar amplitudes
$a \simeq 0.01$ dominates the spectrum. Within this
range, the spectrum contains many formally 
significant periodicities with amplitudes 
$\ge 4 \sigma$ (but none of 
$\ge 5 \sigma$). However, in view of the MOST 2008 
results, it appears that some oscillation power extends
throughout the whole spectrum accessible from the
$V$ data ($0.0003 < f < 0.5$ c/d), but none
was coherent enough to produce a single peak
to be detectable in the ASAS observations.

%Figure: ASAS log-log all seasons  ---------------------------

%--------------------------------------------------------------

\subsection{Fourier analysis of the seasonal data}

In order to find periodic variability during
individual seasons, the ASAS $V$ band data were
analyzed with the seasonal mean brightness
levels subtracted. Such data may give better defined
results than a search for periodicities in the
combined ASAS dataset because 
($i$)~the low frequency, inter-seasonal 
variation is apparently strong and may influence
results for the shorter time scales, and ($ii$)~the MOST 
results suggest short duration of individual
oscillations of only a few oscillation periods.

The seasonal ASAS runs lasted typically 150 -- 250 days
each year (i.e.\ some 3 -- 5 times longer than the 
2008 MOST run) and consisted of 27 to 127
observations. The two last seasons, 2007 and 2008, 
which are most important for a comparison with the MOST data,
were the best observed with 113 and 127 observations,
respectively. The lowest frequencies accessible from the
seasonal datasets are $\simeq 0.005$ c/d. 

\begin{figure}
\begin{center}
\includegraphics[width=80mm]{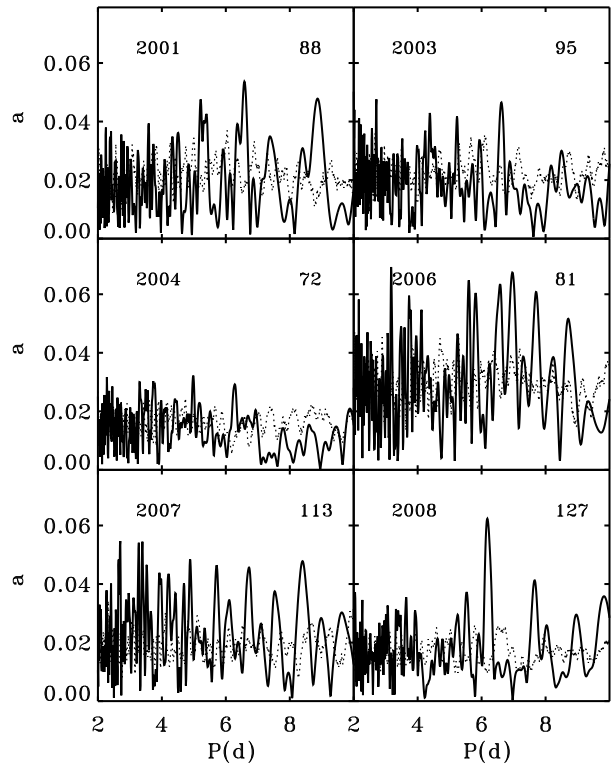}
\caption{The period analysis of the seasonal ASAS data 
in the $V$ band for TW~Hya for the 6 best observed periods,
in the period range of 2 to 10 days only. The 
mean standard error level in each panel is given by the
thin, dotted line. The numbers identify the year and 
give the number of observations.}
\label{ASAS_per}
\end{center}
\end{figure}

The results are presented in the form of 
period -- amplitude spectra
(Figure~\ref{ASAS_per}) for the 6 seasons which had more than
70 observations per season. For the best presentation,
only the period range of 2 -- 10 days is shown and the spectra
are expressed versus the period rather than the frequency. 
This is the range
where very well defined semi-periodic variability was
observed by MOST in 2007 and 2008; we address
the matter of the ASAS observations during the
MOST runs in the next section.

% Figure: ASAS Fourier analysis, 6 seasons, 2001 – 2008 ---

%----------------------------------------------------------

% Figure: ASAS Fourier analysis, 2007 season only --------

\begin{figure}
\begin{center}
\includegraphics[width=80mm]{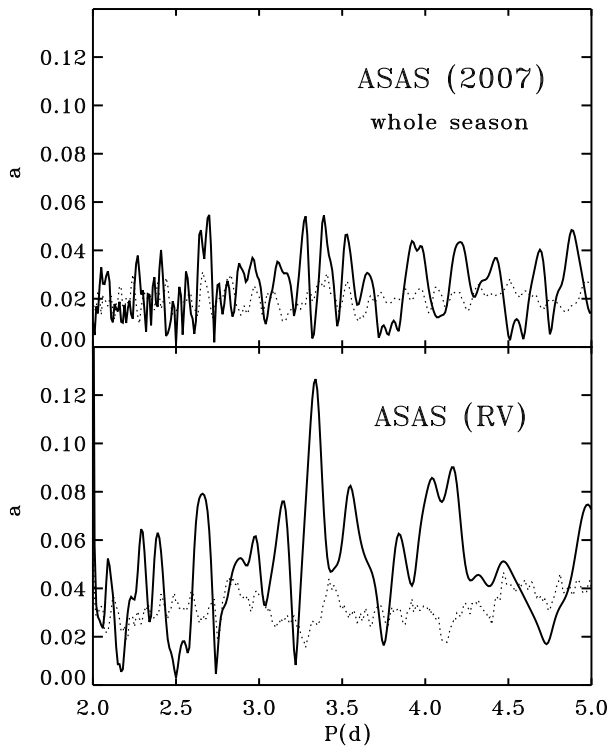}
\caption{The period analysis of the seasonal ASAS data 
in $V$ for TW~Hya for the whole 2007 season in the period
range where the strong periodicity was observed
by MOST is shown in the upper panel. 
The dotted line gives the 
errors of amplitudes. The analysis of the 
reduced set of the ASAS data, obtained within
the duration of the
radial velocity observations of \citet{Set2008}
is shown in the lower panel.}
\label{ASAS_per36}
\end{center}
\end{figure}
%----------------------------------------------------------

\subsection{ASAS observations during the 2007 and
2008 MOST runs}

The comparison of the ASAS and MOST results
cannot be done directly because
the ASAS data extended for typically about 200 days
during each season, while the
MOST runs were comparatively short, of 11 and 46 days.
Also, the ASAS observations were done at intervals
of typically 2 days, sometimes even 3 -- 5 days, 
as can be seen in the distribution of large circles 
among the MOST observations in  
Figures~\ref{MOSTdata7} and \ref{MOSTdata8}. Note that
in these figures, the $V$ band observations were arbitrarily 
adjusted to the assumed mean level of $\bar{V}=11.04$
while the difference in the expected variation amplitudes
between the $V$ band and the MOST band (between $R$ and $I$)
was disregarded. 

We notice in Figure~\ref{ASAS_per} that the 
ASAS 2008 season data show a single, strong, well defined 
periodicity of 6.18 d (0.162 c/d). Surprisingly, it is 
entirely absent in the MOST 2008 spectrum (see
Figure~\ref{MOSTfreq7_1}); the nearest
peaks are at 7.8 and 5.0 days (the second 
ASAS periodicity of 7.67 d may be identical 
with the 7.8 d MOST period). 
This behaviour is very typical for all
ASAS seasons: Some periodic, coherent variations appear 
at various frequencies (usually never the same),
but their amplitudes are not as well defined as in 
the MOST data and they do not directly
show any ``flicker-noise'', $a \propto 1/\sqrt{f}$ 
characteristics. We suspect that the irregular spacing
of the ASAS observations was the reason for the 
latter effect.

A comparison for the 2007 season is most
interesting as during this time the well defined 
periodicities were observed photometrically (MOST, 3.7 d) and
in radial velocities (3.56 d, \citet{Set2008});
within the period uncertainty, this is the same 
variation showing a simple phase relation
between the two observables  
(Section~\ref{same-thing}). Unfortunately, only
five ASAS observations were obtained during the 2007 MOST
run, but they show the perfect
consistency of the ASAS and MOST observations 
(see Figure~\ref{MOSTdata7}). 
However, 45 ASAS observations happened to fall
into the full extent of the radial velocity 
observations \citep{Set2008}. In Figure~\ref{ASAS_per36}, 
we compare the full results for the 2007 season
ASAS observations with the subset obtained during
the time when the radial velocity observations revealed
the very strong 3.56 d periodicity. 
As we can see in Figure~\ref{ASAS_per36}, the longer
ASAS 2007 dataset gives smaller amplitudes, indicating a 
loss of coherency. Very clear periodicities
are visible in the reduced-duration dataset, but none 
corresponds to $\simeq 3.56$ d. However, this may be due to the
irregular data sampling in the ASAS project because
significant periodicities appear at 3.3 d and 4.1 d. 

Thus, the ASAS data are very useful
for an extension of the frequency spectrum to very low 
frequencies beyond the limit of the MOST data (as set
by the duration of the 2008 run), but
the picture obtained for the seasonal ASAS runs is confusing
and would certainly not lead to such clear-cut results
as from the MOST data. The ASAS project,
with its irregular sampling, could not detect single, but
variable periodicities because of the obvious
loss of coherence. But variability is definitely there, 
only one cannot characterize it as well as for the
MOST data.

\section{Summary and discussion}
\label{disc}

Our analysis of TW~Hya contributes substantially 
to the knowledge of T~Tauri type photometric variability. 
Although connection with the stellar rotation cannot be
directly demonstrated, the MOST observations are phenomenologically 
consistent with the Type~IIp variability of TW~Hya,
as suggested by \citet{Herbst1994}. The great
advantage of the MOST observations was their continuous and
uniform time coverage of 11 days in 2007 and 46 days in 2008,
both sampled at the satellite-orbit interval of 0.07 d.
The ASAS observations permitted extension of the 
time-scale coverage to $\simeq 3000$ days but with
irregular and sparse sampling at typically 1 -- 5 days.

During the 2007 MOST observations, a single, strong 
3.7 d period dominated the brightness changes.
This variation had a similar period to the 3.56~d
radial velocity variation with semi-amplitude 
of 200 m/s observed at the same time by \citet{Set2008}.
The phase relation between the two phenomena was such that
the originally offered explanation of an orbiting planet
has great difficulty in explaining the observations:
Only a single (per orbit) brightness maximum 
took place at what -- for the orbital revolution 
hypothesis -- would be interpreted as inferior conjunction
of the planet. While this phase relation could result
from a spot modulation, the presumed low inclination angle 
of the rotation axis makes this hypothesis very unlikely. 
We note that this periodicity 
did not show up directly in the ASAS 
observations during the 2007 season, but the reason for
that may be in the irregular ASAS data sampling.

The 3.56/3.7 d periodic modulation did not re-appear in
the four-times longer MOST run of 2008. All changes during 
this run can be interpreted as semi-periodic events
(mostly brightening spikes) superimposed on a baseline level.
Several Fourier components appeared at periods ranging
between 1.2 and 7.7 d with noticeably smaller 
amplitudes than the 2007 periodicity; they 
had progressively smaller amplitudes with decreasing periods
with an envelope suggesting flicker-noise properties.  
The long MOST 2008 run, analyzed with 
wavelets for the period range $0.2 < P < 20$~d
showed discrete periods subject 
to a well defined period-change pattern: All
variations were mono-periodic, appeared in the 2 -- 9 d range, 
changed their periods, lasted a few cycles 
and then died out. An evolution of the period 
from $\simeq 5 - 6$ d to $\simeq 3$ d in the time 
scale of about 10 d was observed. 
Thus, the Fourier components showing that flicker-noise
amplitude distribution appeared to be another
representation of the mono-periodic features 
evolving within the 2 -- 9 d period range. 
No wavelet-detectable components were seen
for time scales $>9$ days nor for the 
$0.2 < P < 2$ d range, with the latter range being
dominated by small, shot-noise events. 

The variability of TW~Hya is most likely powered by
accretion phenomena in its disk. Because of the
low inclination of the rotation axis, the disk 
is visible almost face on with its inner parts exposed. 
The star inside the disk is relatively small; it has the
radius $\simeq 1\,R_\odot$, i.e.\ only slightly larger 
than a Main Sequence (MS) star of the same spectral type. 
This can be estimated using 
the Hipparcos distance $d=56$ pc and the
faintest level ever observed, $V \simeq 11.3$ 
\citep{RK1983}, hence the luminosity of the ``naked''
star is expected to be $M_V \simeq 7.56$. 
For comparison, a Main Sequence K5V star would have 
$M_V \simeq 7.8$, while a K7V star would have 
$M_V \simeq 8.1$. Thus, the small over-luminosity may be 
interpreted by a radius excess of 12 to 28 percent over
the MS value. 

The Keplerian orbits for a $0.72\,M_\odot$ star
(for consistency we assume the same mass as in \citet{Set2008})
with periods 2, 4, and 8 days would be located at the disk
radii of 6, 9.5, and 15 $R_\odot$. These may be 
the locations where bright, hot patches or blobs
of changing optical depth formed and 
remained stable for a few orbital revolutions. Although we
do not know the exact mechanism, and we cannot
disentangle separate contributions from the
changing area, optical depth or temperature, we
suspect that the latter dominates 
because of the strong brightness --
colour correlation in the variations observed
before \citep{RK1983}. TW~Hya definitely
requires a well organized observational effort,
involving spectroscopy supporting further MOST
satellite observations.   

Big questions remain: 
The rotation period of TW~Hya is still unknown; the previously
suggested value of about 2 days is just a plausible guess. 
We also have no interpretation or mechanism
for the strong 2007 periodicity. The brightness variations
with the shortening periods observed in 2008 are
unexplained but is is virtually impossible
to explain all these phenomena by surface spots.
Large, optically thick, hot-plasma structures, 
anchored in different parts of the inner accretion disk 
would be better candidates. They could
extend to large latitudes and explain the brightness 
variations through changes in their optical depth even 
for a low rotational inclination angle.

\section{Conclusions}
\label{concl}

Following our analysis of the MOST and ASAS very different
but partly complementary 
observational data sets, and considering conclusions of
the important paper by
\citet{Press1978} on the common occurrence
of flicker noise (with Fourier amplitudes scaling 
as $a \propto 1/\sqrt{f}$), we cannot resist a comparison
of the TW~Hya photometric variability with an orchestra. 
As pointed out by Press, the
spectrum of the orchestra sound can be
usually described by the prosaic flicker noise. 
Thus, although we hear the music defined by an  
entire orchestra where everybody plays a different piece,  
sometimes the melody (tune) produced by an 
individual instrument becomes noticeable
against the general flicker-noise level. 
We observed such distinct, time-variable ``tones''
in the period range of 2 to 9 days; they may be
manifestations of the accretion process as the matter
spirals into the star. 

The flicker-noise spectrum of TW~Hya photometric 
variations appears to extend from very 
low frequency of 0.0003 c/d (accessible from 8 years of 
the ASAS data) to $\simeq 10$ c/d
(accessible to our MOST observations). The lowest frequencies
require long monitoring time, so that good photometric
stability and consistency of calibrations are essential, 
even at relatively moderate accuracy (0.01 -- 0.02 mag)
of the ASAS project. At frequencies corresponding to days
to weeks, the uniform time coverage and high accuracy
of the MOST mission (0.003 -- 0.005 mag, even with 
the included intrinsic variability of the star in 
times scales shorter than an hour) permitted us to
study the difficult (from the ground) frequency range
of $\simeq 0.025$ c/d to 10 c/d. In this range, we observed
very clear, well defined, varying ``tones'' 
in the 0.1 -- 1.0 c/d range only (specifically,
the periods 2 -- 9 days). It would be tempting
to identify those changing periods as signatures
of the orbital decay as the accreting material
spirals into the star.
The strong periodicity observed by MOST in 2007 with
the period of 3.7 d, in phase with the 3.56 d 
radial-velocity variations of \citet{Set2008}, appears to
be one such ``tone''. It does not seem to be 
related to a possible planet orbiting 
TW~Hya because it disappeared
within one year. However -- contrary to the
clear period changes observed in 1.5 month of the MOST
observations in 2008 -- it was relatively stable
through the 3 month of the 2007 radial velocity observations. 

The wide range of variability frequencies suggests
a multitude of mechanisms. While we suspect that the main
mechanism in the range of time scales accessible to MOST 
is accretion within the innermost disk at distances 
of 2 -- 15 $R_\odot$ from the star, it is hard to imagine
that accretion would produce photometric variability 
on time scales of years at the implied radii of 
several astronomical units. 
Thus, in the symphony orchestra analogy,
more instruments (mechanisms) must be contributing to
create the extended, well defined flicker-noise 
variability spectrum of TW~Hya.

\section*{Acknowledgments}
The Natural Sciences and Engineering Research Council of
Canada supports the research of DBG, JMM, AFJM, and SMR.
Additional support for AFJM comes from FQRNT (Qu\'ebec). 
RK is supported by the Canadian
Space Agency and WWW is supported by the Austrian Space
Agency and the Austrian Science Fund. 

Special thanks are extended to Ray Jayawardhana,  
Marten van Kerkwijk and Alexis Brandeker
for very useful comments and to John Percy 
for an excellent review. 

This research has made use of the SIMBAD database,
operated at CDS, Strasbourg, France and NASA's Astrophysics 
Data System (ADS) Bibliographic Services.

\end{document}